\documentstyle[multicol,aps,epsf,here]{revtex} 

\begin{document}
\draft

\title{Gold Nanowires and their Chemical  Modifications}

\author{
Hannu H\"akkinen, Robert N. Barnett, and Uzi Landman}
\address{School of Physics, Georgia Institute of Technology, 
Atlanta, GA 30332-0430}
\date{August 1999}
\maketitle

\begin{abstract}
Equilibrium structure, local densities of states, and electronic 
transport in a gold nanowire made of a four-atom chain
supported by two gold electrodes, which has been imaged 
recently by high-resolution electron microscopy, and
chemical modification of the wire via the adsorption
of a methylthiol molecule, are investigated with ab-initio
local density functional simulations. In the bare wire
at the imaged geometry the middle two atoms dimerize, and the structure
is strongly modified by the adsorption of the molecule with
an accompanying increase of the ballistic conductance
through the wire. 

\pacs{PACS: 73.40.Jn, 73.61.-r, 85.30.Vw }
\end{abstract}

\begin{multicols}{2}
\narrowtext

Generation of wires of atomic scale dimensions
in the process of formation and/or elongation
of interfacial contacts has been predicted through
early molecular dynamics simulations using many-body
potentials \cite{Lan90,Lan91}, and in the face of fundamental
interest and technological considerations driven by
the relentless miniaturization of electronic and
mechanical devices such wires have been the subject of
intensive experimental and theoretical research
endeavors \cite{Ser97}. Indeed, nanometer-scale wires
(nanowires, NWs) have been created and their structural,
mechanical, and transport characteristics were studied
using a variety of techniques \cite{Ser97},
including most recently combined scanning tunneling
microscopy (STM) and direct imaging with the use of
high-resolution transmission electron microscopy
(HRTEM) \cite{Kiz97,Kiz98,Ohn98}.

We report here on ab-initio local-density functional
(LDA) investigations \cite{Note1} of the
atomic structure, electronic spectrum and conductance
of a gold NW consisting of a four-atom chain
connected to gold electrodes, which is the smallest
NW imaged by HRTEM \cite{Ohn98,Note2}.
Our study reveals dimerization of the gold atoms
in the middle of the chain \cite{Note2}
akin to a Peierls transition in (extended) one-dimensional
systems. Furthermore, we explore structural
and electronic spectral modifications resulting from
adsorption of a molecule (methylthiol, SCH$_3$)
to the wire \cite{Li98},  demonstrating the sensitivity of
these properties to such chemical interactions,
 as well as their effect on   the electronic conductance
of the wire which we find to increase upon adsorption.
These results provide a new interpretation of 
the measured HRTEM image of the atomic gold
wire \cite{Ohn98,Note2} and suggest a new strategy for formation
of organo-metallic NWs, as well as the use of NWs
as monitoring and chemical sensing devices. 
 
In light of the aforementioned STM/HRTEM observations
\cite{Ohn98,Note2} we start our simulation from
a 4-atom Au wire consisting of two tip-atoms (t)
located at the apexes of two opposing tip-electrodes
distanced initially by $d_{tt}^{(0)}$=8.9 \AA, and of two
internal Au atoms (i) located in the gap between
the two electrodes with initial uniform distances
$d_{ti}^{(0)}$=$d_{ii}^{(0)}$=2.967 \AA\ between
neighboring  atoms of the wire; the tip-electrodes 
consist each of 29 gold atoms arranged in a 
pyramidal shape made of face-centered-cubic 
stacked (110) layers and exposing
(100) and (111) facets, with the tip atoms (as well
as the internal wire atoms) and the atoms of the underlying
layers supporting them treated dynamically, while the rest 
of the electrode gold atoms are held
at their crystalline lattice positions. This initial structure 
relaxes spontaneously in the course of 
a total energy minimization with a 0.16 eV gain in
the total energy, to that shown in 
Fig. 1 (AuNW, left) where  the two inner wire atoms
dimerize, with $d_{ii}$=2.68 \AA\ (compared to
$d$(Au-Au)=2.48 \AA\ in a  free Au$_2$ molecule),
$d_{ti}$=3.07 \AA,  and the total length of the wire
shortens to $d_{tt}$=8.82 \AA\ (see ref. \onlinecite{Note3}).

From the local density of states (LDOS) of the dimerized
wire (Fig. 2), calculated in the regions delineated 
in Fig. 1 (AuNW, left), we observe that the electronic states in the 
interior region of the wire (region E, see Fig. 1, left)
near the Fermi energy ($E_F$, marked by a dashed line
in Fig. 2) are not found in the free Au$_2$ dimer (whose
states are marked by dots on the upper energy axis of Fig. 2).
Rather, these states originate   from hybridization
between the atomic states of the interior Au atoms
and the gold-electrode states, with the highest-occupied
molecular orbital (HOMO) exhibiting a dominant
d-character on the interior wire atoms.
The dimerization is driven by lowering 
of the energies of states in the interval
$-10\leq E\leq -6.5$ eV calculated at the interior
region of the wire (compare the upper and
and lower LDOS curves in the inset to Fig. 2,
corresponding to the dimerized and initial
equal-distance configurations, respectively).
Additionally, the dimerization of the wire 
is accompanied by a small increase of the
energy gap near $E_F$ from 0.194 eV in the
equidistant wire to 0.216 eV in the dimerized one;
in a certain sense the observed dimerization
may be regarded as a "predecessor" of a
Peierls transition, though such description
should be treated with caution due to the 
rather limited extent of the wire considered
here.  The calculated electronic conductance
\cite{Note4} of the dimerized AuNW is 
$G$=0.58 $g_0$ ($g_0$=$2e^2/h$, where $e$ is the electron 
charge and $h$ is the Planck constant)
corresponding to a resistance of 22.17 k$\Omega$,
and ballistic transport occurs through 
a single conductance channel \cite{Note4}.

Two binding configurations of a methylthiol
molecule to the dimerized equilibrium
configuration of the AuNW were considered:
(i) the SCH$_3$ molecule bonded to the two
middle interior Au atoms of the wire (see
AuNW/m-SCH$_3$ in Fig. 1), and (ii)
binding of the molecule to a terminal tip
atom (t) of the wire and to the neighboring
interior wire atom (see AuNW/t-SCH$_3$ in
Fig. 1). The binding energies of the molecule in
the two adsorption configurations are
essentially the same (4.01 eV), and in both
cases binding of the molecule is accompanied
by significant structural changes of the wire.
In the m-SCH$_3$ configuration (Fig. 1, middle)
the dimerization of the interior wire gold atoms
is removed, and 
the sulfur atom is bonded to the two interior
gold atoms with 
$d$(S-Au)=2.31 \AA,  the angle
$\angle$(Au-S-Au)=117.4$^o$, and $d$(S-C)=1.84 \AA. The
configuration is symmetric about a plane
of reflection passing through the sulfur atom
normal to the plane of the figure, and the length
of the wire increases by 0.1 \AA\ (i.e.,
$d_{tt}$=9.02 \AA), with $d_{ti}$=2.58 \AA\
and $d_{ii}$=3.94 \AA; the wire atoms
are shifted slightly in  the direction
of the adsorbed molecule with the terminal 
wire atoms displaced laterally  by
0.016 \AA\ from the four-fold
hollow site of the 
underlying gold electrode layer.
Dimerization of the interior gold wire atoms
is removed also 
in the t-SCH$_3$ equilibrium bonding configuration
where symmetry is broken, the bond lengths of the
sulfur to the two gold atoms are unequal
[$d$(S-Au(t))=2.38 \AA\ and
$d$(S-Au(i))=2.31 \AA], the angle
$\angle$(Au(t)-S-Au(i))=109.3$^o$,
$d_{ti}$=3.83 \AA, $d$(S-C)=1.85 \AA,
and the length of the
wire is $d_{tt}$=8.98 \AA; additionally,
the Au(t) atom bonded to the sulfur is
displaced laterally by 0.057 \AA\ from the
four-fold hollow site, while the displacement 
of the other Au(t) atom  (not bonded directly
to the molecule) is 0.012 \AA.
We remark that in both adsorption configurations
the intra-thiol $d$(S-C) distance, as well as
the S-Au bond length are similar to those
calculated for the equilibrium structure of a 
free Au$_2$SCH$_3$ molecule where 
$d$(S-C)=1.84 \AA\ and $d$(S-Au)=2.41 \AA,
while the Au-S-Au angles in the thiolated wires are
much larger than in the free molecule
(where $\angle$(Au-S-Au)=67.1$^o$) due to
the binding of the sulfur-bonded gold atoms
of the wire to the rest of the nanostructure.

Examination of the LDOS for the chemically modified
wires displayed in Fig. 3 reveals that changes from the
spectrum of the bare dimerized wire (Fig. 2) are localized to
regions in the immediate vicinity of the molecular 
binding site (compare region E in Fig. 2
with region T for AuNW/m-SCH$_3$ and regions
T and E' for AuNW/t-SCH$_3$ in Fig. 3). In both
cases $E_F$ lies below the HOMO level of the free
SCH$_3$ molecule correlating with energy gain
due to hybridization of the molecular states
with the states of the wire. In both of the
chemically modified wires the HOMO level is 
partially occupied (containing about one hole)
while it is doubly occupied in the bare AuNW,
and the gap between that level and the lowest
unoccupied one is 0.21 eV and 0.316 eV in the
m-SCH$_3$ and t-SCH$_3$, respectively.

Chemical modifications of the AuNW are
signaled, and may be detected, by changes
in the electronic conductance, which increases
upon adsorption, i.e. $G$(AuNW/m-SCH$_3$)=0.82
$g_0$ and $G$(AuNW/t-SCH$_3$)=0.88 $g_0$, 
involving a single conductance channel.
While at first sight an increase of the conductance
in the presence of an adsorbate may seem surprising,
it can be explained via examination of the potential
landscapes governing the propagation of the electron
through the wires (see Fig. 4).  Comparison 
between the potential shown in Fig. 4a for the
bare dimerized nanowire with those corresponding to the
chemically modified ones (Fig. 4b and 4c),
reveals that the potential barriers (bottle-necks),
associated with the unequal spacings between
the gold atoms in the dimerized equilibrium
structure and the reduced overlap between
the electronic states of the inner wire and the
tip atoms, which decrease the transmission of the
incident electron through the bare AuNW (Fig. 4a),
are reduced in the chemically modified wires and the
conductance path is broadened, resulting in
enhancement of the ballistic transmission
through these wires \cite{Note5}.

These findings, obtained through ab-initio simulations,
pertaining to the dimerized structure of an atomic
gold nanowire and the sensitivity of structural,
electronic, and conductance properties of such wires
to chemical modifications via molecular adsorption,
provide a new interpretation of recent HRTEM measurements
\cite{Ohn98,Note2},  demonstrate methods 
for probing the nature of chemical interactions
with such nanostructures, and suggest a strategy
for preparation of chemically modified nanowires.

\bigskip

{\it Acknowledgement.} 
We thank A.G. Scherbakov for his assistance in 
conductance calculations. 
This research is supported by the U.S. DOE,
AFOSR, and the Academy of Finland. 
Calculations were performed on an
IBM SP2 parallel computer at the Georgia Tech
Center for Computational Materials Science,
and on a Cray T3E at the National Energy Research Scientific
Computing Center (NERSC) at Berkeley,
CA.

\begin{figure}
\caption{
Equilibrium structures of a bare gold nanowire
(AuNW, left) and of wires chemically modified
by adsorption of a SCH$_3$ molecule, with 
the molecule adsorbed in the middle of the wire
(m-SCH$_3$) or at the vicinity of the tip
(t-SCH$_3$). The bare 4-atom wire consists of
two tip atoms (t) and two interior atoms
(i), and it is supported by two opposing fcc
gold electrodes exposing (100) and (111)
facets, with the wire axis along the [110]
direction. Yellow spheres correspond to 
Au atoms, and in the chemically modified wires
S, C, and H atoms are depicted by red, green, and blue
spheres, respectively. The length of the
dimerized bare AuNW $d_{tt}$=8.82 \AA, and those
of the AuNW/m-SCH$_3$ and AuNW/t-SCH$_3$ wires
are 9.02 \AA\ and 8.98 \AA, respectively.
Regions in the wires used for LDOS calculations
(see Figs. 2 and 3) are denoted by letters.
Marked distancies are in units of \AA.
}
\end{figure}

\begin{figure}
\caption{
Local densities of states (LDOS) for the dimerized AuNW.
Different curves are marked by letters corresponding to
the regions delineated in Fig. 1 (left),
and they are vertically shifted with respect
to each other for clarity. Displayed in the inset
are the LDOS calculated in the interior region
(E) of the dimerized wire (upper curve) and the initial
equidistant wire (lower curve). The Fermi energy
($E_F$=--5.68 eV) is denoted by a dashed line. The
molecular eigenvalues of a free gold dimer (Au$_2$)
are marked by dots on the upper axis (filled and empty
dots correspond to occupied and unoccupied states,
respectively). Energy in units of eV and LDOS in 
eV$^{-1}$/atom.
}
\end{figure}

\begin{figure}
\caption{
LDOS for chemically modified AuNW, with a SCH$_3$
molecule adsorbed at the middle of the wire
(AuNW/m-SCH$_3$, left) and at the tip-vicinity
(AuNW/t-SCH$_3$, right).  Different curves
marked by letters correspond to regions delineated in
Fig. 1. In regions T, which include the adsorbed thiol
molecule, occupied states of the free SCH$_3$
molecule are denoted by filled dots on the upper
axis. The Fermi energy is denoted by a dashed
line; $E_F$=--5.59 eV and --5.66 eV for the 
m-SCH$_3$ and t-SCH$_3$ adsorption configurations.
Energy in units of eV and LDOS in 
eV$^{-1}$/atom.
}
\end{figure}

\begin{figure}
\caption{
Potential profiles [11] in cross sectional
cuts through the bare AuNW dimerized wire
(a) and through the chemically modified ones (b and c).
The vacuum level is at zero and the potential step
between adjacent contours is 0.05 eV. Note
the potential bottle-necks for the bare dimerized wire
(a), which are highly reduced in the chemically 
modified nanowires. Energy in units of eV
and distances in Bohr radius ($a_0$).
}
\end{figure}

\end{multicols}
\end{document}